\documentclass[letterpaper]{IEEEtran}
\usepackage{cite} 
\usepackage{graphicx,subfigure}
\usepackage[justification=justified]{caption}
\usepackage{algpseudocode}
\usepackage{setspace}
\usepackage{epstopdf}
\usepackage{tikz}
\usepackage[absolute]{textpos}
\usepackage{xcolor}
\usepackage{xspace,bm}
\usepackage{amsmath}
\usepackage{amsfonts}
\usepackage{epsfig}
\usepackage{amssymb}
\usepackage [english]{babel}
\usepackage [autostyle, english = american]{csquotes}
\MakeOuterQuote{"}

\usepackage[T1]{fontenc}
\usepackage[utf8]{inputenc}
\usepackage{authblk}

\newcommand{\remove}[1]{}

\title{\Huge Fault Adaptive Routing in Metasurface Controller Networks \\
\large (Invited Paper)}

\author[1,2]{Taqwa Saeed}
\author[2]{Constantinos Skitsas}
\author[2]{Dimitrios Kouzapas}
\author[1]{Marios Lestas}
\author[3]{Vassos Soteriou}
\author[2]{Anna Philippou}
\author[4]{Sergi Abadal}
\author[5]{Christos Liaskos}
\author[6]{Loukas Petrou}
\author[6]{Julius Georgiou}
\author[2]{Andreas Pitsillides \vspace{-0.2cm}}

\affil[1]{Department of Electrical Engineering, Frederick University Cyprus, Nicosia, Cyprus}
\affil[2]{Department of Computer Science, University of Cyprus, Nicosia, Cyprus}
\affil[3]{Dept. of Electrical Eng., Computer Eng. and Informatics, Cyprus University of Technology, Limassol, Cyprus}
\affil[4]{Department of Computer Architecture, Universitat Politècnica de Catalunya, Barcelona, Spain}
\affil[5]{Institute of Computer Science, Foundation of Research and Technology Hellas, Heraklion, Greece}
\affil[6]{Department of Electrical and Computer Engineering, University of Cyprus, Nicosia, Cyprus}
\affil[ ]{Emails:\{st009698@stud.fit.,eng.lm@frederick\}.ac.cy$^1$,\{cskits01,dkouza01,annap,Andreas.Pitsillides\}@cs.ucy.ac.cy$^2$, \{vassos.soteriou@cut.ac.cy\}$^3$,\{abadal@ac.upc.edu\}$^4$,\{cliaskos@ics.forth.gr\}$^5$,\{petrou.loukas,Julio\}@ucy.ac.cy$^6$ \vspace{-0.5cm}}

\TPMargin{5pt}
\newcommand{\copyrightstatement}{
    \begin{textblock}{0.41}(0.08,0.95)   
	\textblockcolour{white}
         \noindent
         \footnotesize
          978-1-5386-8552-5/18/\$31.00~\copyright 2018 IEEE \hfill
    \end{textblock}
}

\begin{document}

\copyrightstatement
\maketitle \thispagestyle{empty} \pagestyle{empty}

\begin{abstract}
HyperSurfaces are a merge of structurally reconfigurable metasurfaces whose electromagnetic properties can be changed via a software interface, using an embedded miniaturized network of controllers, thus enabling novel capabilities in wireless communications. Resource constraints associated with the development of a hardware testbed of this breakthrough technology necessitate network controller architectures different from traditional regular Network-on-Chip architectures. The Manhattan-like topology chosen to realize the controller network in the testbed under development is irregular, with restricted local path selection options, operating in an asynchronous fashion. These characteristics render traditional fault-tolerant routing mechanisms inadequate. In this paper, we present work in progress towards the development of fault-tolerant routing mechanisms for the chosen architecture. We present two XY-based approaches which have been developed aiming to offer reliable data delivery in the presence of faults. The first approach aims to avoid loops while the second one attempts to maximize the success delivery probabilities. Their effectiveness is demonstrated via simulations conducted on a custom developed simulator.  
\end{abstract}
\vspace{-0.4cm}
\let\thefootnote\relax\footnote{This work was partially funded by the European Union via the Horizon 2020: Future Emerging Topics
call (FETOPEN), grant EU736876, project VISORSURF
(http://www.visorsurf.eu) and Cyprus RPF HSadapt, COMPLEMENTARY/0916/0008.}
\section{Introduction}
\label{intro}

Hypersurfaces (HSFs) is a recently proposed \cite{liaskos2015design} metamaterial based paradigm, envisioned to realize a new generation of applications, as for example programmatically controlled wireless environments \cite{liaskos2018new}. The core technology behind the paradigm is metasurfaces \cite{yang2016programmable}, which are planar artificial structures comprising a periodically repeated element, the meta-atom, over a substrate. Metasurfaces, may be engineered to possess customized electromagnetic (EM) characteristics, fully defined by the chosen form of the meta-atom. Such EM characteristics can be used to realize application related functionalities such as perfect absorption, beam steering via anomalous reflection, or polarization control.

Early metasurface structures were static in nature, severely limiting their scope. To this end intense research activity has been reported on reconfigurable metamaterials and metasurfaces \cite{abadal2017computing}.
Although dynamic metasurfaces are a major breakthrough, the lack of programmatic control over the functionality has motivated the introduction of the concept of software defined metasurfaces or HSFs.       

The main underlying idea is the introduction of a network of miniaturized controllers, through which software directives are transformed into reconfiguration stimuli on the metasurface, resulting to changes in the meta-atom structure and thus the EM properties of the metasurface \cite{petrou2018asynchronous}. In the simplest version of this concept, the controllers activate or deactivate corresponding switches. 
The controller network (CN) is connected to external network devices via a gateway. The CN design is challenged by a number of factors\cite{kouvarosformal}, as for example the small meta-atom size, the possibly large number of nodes that need to be accommodated, the need to avoid EM interference and the presence of faults. The above necessitate simple, low cost, power constraints and fault tolerant solutions.  

The system-level resemblance between multiprocessors and the HSF CN indicates that the Network-on-Chip (NoC) methodologies can be adopted to the HSF paradigm \cite{abadal2017computing, vangal200880}. However, the shift in the design requirements and primarily the need for simplicity suggest that although NoC methodologies can serve as starting points in the design procedure, custom solutions need to be developed. A major objective of our current work is the development of a prototype to realize the HSF concept. Taking into consideration the design specifications outlined above and additional hardware constraints, a CN architecture has been adopted which is characterized by asynchronous operation, an irregular Manhattan-like topology and a single gateway (GW) with a single point of entry coupled with a single acknowledgment gateway (ACK GW).            

Aiming towards the sustainment of dependable communication among interconnected controllers in an HSF structure, the introduction of link-level fault tolerance in its underlying CN, where faulty links are bypassed by control messages using a suitably designed fault-tolerant (FT) routing function, becomes paramount. 
FT approaches applicable to NoCs have been inspired from macro-level interconnection networks, where Radetzki’s survey \cite{radetzki2013methods} provides a broad coverage in the field. Essentially, as long as FT routing provides full connectivity devoid of cyclic channel dependencies in a sub-connected (i.e., faulty) topology, then the FT function crucially guarantees deadlock- and livelock-freedom during packet delivery. Such deadlock-avoidance mechanisms are explored in techniques such as in the Turn Model for adaptive routing \cite{glass1992turn}, where certain 90 degree turns in mesh interconnection networks are prohibited so as to break the formation of cycles and channel dependencies. 

Many FT routing algorithms build on the principles above to ensure seamless communication in NoCs, such as the FT scheme in \cite{vitkovskiy2013dynamic} which creates a local detour every time a faulty node is encountered. As such, new path directives are added to the header which can create a large overhead. The method proposed in \cite{aisopos2011ariadne}, on the other hand, works proactively by disseminating routing data that is used to update routing tables right after a new fault is encountered, so as to bypass faulty links and routers. Another approach in \cite{wu2003fault} where the author employs the notion of faulty blocks, where healthy links are victimized along with spatially close faulty links, to design a deterministic FT method where routed packets bypass such faulty blocks. Last, in an attempt to sustain good performance while avoiding deadlocks, several authors have utilized hardware-expensive virtual channels \cite{radetzki2013methods}. These concepts, however, are costly in terms of buffering, protocols, and control resources, which cannot be afforded in the HSF CN.

As such, here we seek to develop a lightweight fault adaptive routing protocol which is simple, yet effective, which suits the hardware investment restrictions imposed upon the HSF CN (identified in \cite{kouvarosformal, petrou2018asynchronous}), such as by involving asynchronicity in communication among controllers. Based on the XY routing, primarily proposed for NoC, we introduce two fault adaptive routing algorithms. The one is based on alternating between XY and YX routing in order to tolerate faults detected online, and employs turn prevention to avoid cycles. The other algorithm aims to achieve high reliability by detecting two disjoint paths between any two nodes in the network, such that when a fault occurs in a path, the packet is forced to take an alternative path that does not overlap with the faulty one. The effectiveness of the methods is demonstrated using simulations at this stage. Testing on the prototype to be developed is scheduled in the near future. 
Both methods however, will be further improved and evaluated in the presence of more sophisticated fault models.  

The rest of the paper is organized as follows. Section \ref{Arch} discusses the considered network architecture, in Section \ref{algorithms} the proposed routing algorithms are introduced, in Section \ref{SimEx} the simulation results are discussed and finally concluding remarks and future plans are summarized in Section \ref{conc}.
\begin{figure}[t]
	\centering
		\includegraphics[width=9 cm]{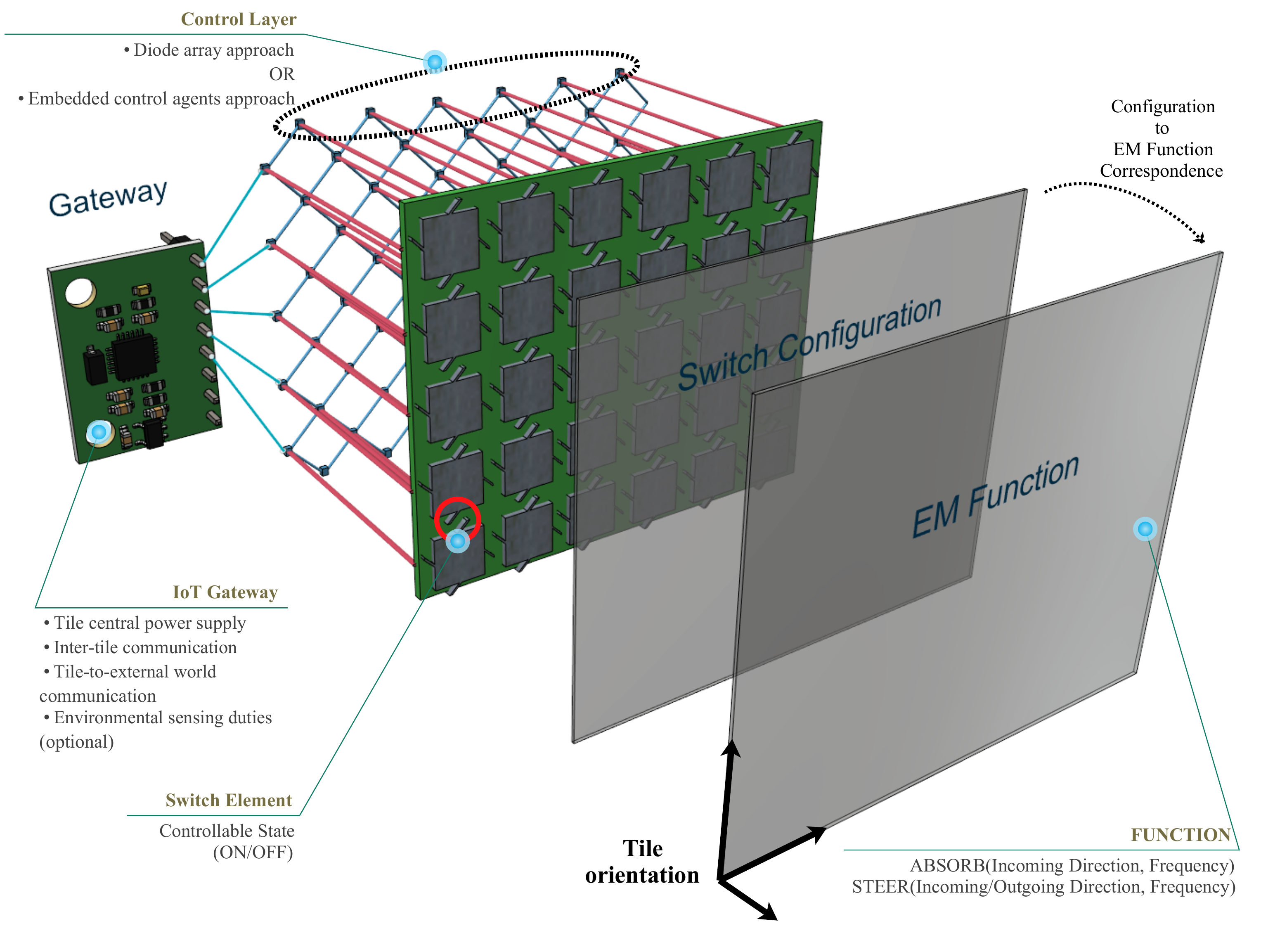}
	\vspace{-0.2cm}
	\caption{The layout of the HSF system.}
	\vspace{-0.2cm}
	\label{HSFF}
\end{figure}
\vspace{-0.4cm}
\section{HyperSurface Controller Network}
\label{Arch}
A schematic overview of the HSF architecture is shown in Fig. \ref{HSFF} \cite{liaskos2018new}. It comprises of a network of miniaturized controllers, each of which controls one or more metasurface switches. These switches, depending on their status, dictate different switch configurations which in turn change the meta-atom structure, thus modifying the electromagnetic functions of the system. The gateway, through a master-slave behavior, provides connectivity between the CN and external networking devices. Although a single tile of controllers is shown in Fig. \ref{HSFF}, provisioning is made to allow for multiple tiles, with the gateway also offering inter-tile communication. 

A first HSF prototype is targeted in this work \cite{petrou2018asynchronous}. Towards the development of this prototype a three layer PCB structure has been considered where the top layer consists of the metal patches, the second layer consists of the ground plane and the bottom layer consists of an array of applications specific integrated circuit (ASIC) in which the controller functionality is embedded. The bottom layer is connected to the top metal patch layer through vias. The ASIC role is to adapt the EM properties of the top layer by providing adjustable complex impedance loading as well as networking functionality. The adjustable complex impedance loading is offered by digitally controlled varactors and varistors. A number of system requirements and constraints dictate the CN architecture design. These are reviewed below.        
\vspace{-0.3cm}
\subsection{Design Specifications}
The system must be characterized by simplicity of implementation, low power consumption and low cost. These are dictated by the  need for scalability, due to the large number of meta-atoms that will need to be accommodated in practice, the small meta-atom size required for correct metamaterial operation at small wavelengths and the need to avoid EM interference. The meta-atom size is of critical importance to the ability of the metasurface to control EM waves. The size of the meta-atom should be comparable to the incident wave wavelength $\lambda$ (of the order of $\lambda$ /2), while the metasurface thickness should be much smaller than the wavelength (of the order of $\lambda$ /10). Taking into consideration that at least 5 meta-atoms per wavelength are required for correct operation,  to accommodate for example 60GHz communications, sizes less than 1mm$\times$1mm are required. Such small meta-atom sizes have led to considering a single chip ASIC for each meta-atom. In addition, the small meta-atom size indicates that a large number of meta-atoms and thus controllers will be involved in practical applications. This implies that the cost of each tile must be kept at a minimum and that the pursued solutions must be able to scale well with the network size. The potentially large surfaces to be covered also necessitate for designs with low power consumption. 

The need to avoid EM interference with the incident waves is also of utmost importance. A large surface which is clocked can potentially radiate significant interference signals and thus prompts for an asynchronous digital design. Asynchronous design is also preferable in terms of scalability and cost as there will be no need for oscillators within a dense array, thus reducing space and cost. Moreover asynchronous circuits are considered to be extremely energy efficient. Electromagnetic interference poses constraints on the network wiring and thus the topology. Wiring should be kept at a minimum, thus favoring a grid networked controller approach. 
                         
Another significant factor to be considered is the issue of fault tolerance. The CN architecture must offer reliable data delivery even in the case of faults, which can be expected due to component failure, external influence and loss of connectivity. It must be noted, however, that different from the traditional NoC stringent performance requirements, HSF applications are expected to have somewhat relaxed requirements in terms of latency and reliability. This implies that failure of some software reconfiguration directives to be successfully delivered to the controller nodes might not be observable at the macroscopic level. Another factor to be accounted for, is that the workload that most applications are expected to incur is relatively low. So summarizing, the CN will be characterized by asynchronous operation, simplicity of implementation, fault tolerance, low power consumption and grid connectivity.         
\subsection{Controller Network Topology}
\label{Archsub}
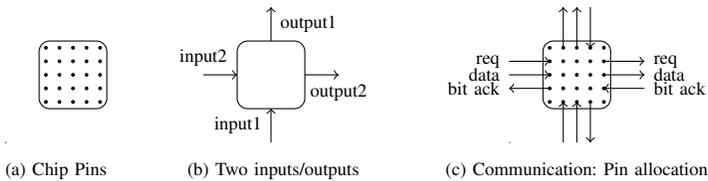
\begin{figure}
	\begin{center}
	\begin{tabular}{ccccc}
		\begin{tikzpicture}[scale=0.9]
			\draw		(0, 0) -- (0, 0);
			\draw[rounded corners]		(0.5, 0.5) rectangle (1.5, 1.5);
			\draw[fill]	(0.6, 0.6) circle (0.02);
			\draw[fill]	(0.8, 0.6) circle (0.02);
			\draw[fill]	(1, 0.6) circle (0.02);
			\draw[fill]	(1.2, 0.6) circle (0.02);
			\draw[fill]	(1.4, 0.6) circle (0.02);

			\draw[fill]	(0.6, 0.8) circle (0.02);
			\draw[fill]	(0.8, 0.8) circle (0.02);
			\draw[fill]	(1, 0.8) circle (0.02);
			\draw[fill]	(1.2, 0.8) circle (0.02);
			\draw[fill]	(1.4, 0.8) circle (0.02);

			\draw[fill]	(0.6, 1) circle (0.02);
			\draw[fill]	(0.8, 1) circle (0.02);
			\draw[fill]	(1, 1) circle (0.02);
			\draw[fill]	(1.2, 1) circle (0.02);
			\draw[fill]	(1.4, 1) circle (0.02);

			\draw[fill]	(0.6, 1.2) circle (0.02);
			\draw[fill]	(0.8, 1.2) circle (0.02);
			\draw[fill]	(1, 1.2) circle (0.02);
			\draw[fill]	(1.2, 1.2) circle (0.02);
			\draw[fill]	(1.4, 1.2) circle (0.02);

			\draw[fill]	(0.6, 1.4) circle (0.02);
			\draw[fill]	(0.8, 1.4) circle (0.02);
			\draw[fill]	(1, 1.4) circle (0.02);
			\draw[fill]	(1.2, 1.4) circle (0.02);
			\draw[fill]	(1.4, 1.4) circle (0.02);
		\end{tikzpicture}
		&\qquad&
		\begin{tikzpicture}[scale=0.9]
			\draw[rounded corners]		(0.5, 0.5) rectangle (1.5, 1.5);
			\draw[->]	(0, 1) node[above] {\scriptsize input2} -- (0.5, 1);
			\draw[->]	(1.5, 1) -- (2, 1)  node[below] {\scriptsize output2} ;
			\draw[->]	(1, 0)  -- node[left] {\scriptsize input1} (1, 0.5);
			\draw[->]	(1, 1.5) -- node[right] {\scriptsize output1} (1, 2);
		\end{tikzpicture}
		&\qquad&
		\begin{tikzpicture}[scale=0.9]
			\draw		(0, 0) -- (0, 0);
			\draw[rounded corners]		(0.5, 0.5) rectangle (1.5, 1.5);
			\draw[fill]	(0.6, 0.6) circle (0.02);
			\draw[fill]	(0.8, 0.6) circle (0.02);
			\draw[fill]	(1, 0.6) circle (0.02);
			\draw[fill]	(1.2, 0.6) circle (0.02);
			\draw[fill]	(1.4, 0.6) circle (0.02);

			\draw[fill]	(0.6, 0.8) circle (0.02);
			\draw[fill]	(0.8, 0.8) circle (0.02);
			\draw[fill]	(1, 0.8) circle (0.02);
			\draw[fill]	(1.2, 0.8) circle (0.02);
			\draw[fill]	(1.4, 0.8) circle (0.02);

			\draw[fill]	(0.6, 1) circle (0.02);
			\draw[fill]	(0.8, 1) circle (0.02);
			\draw[fill]	(1, 1) circle (0.02);
			\draw[fill]	(1.2, 1) circle (0.02);
			\draw[fill]	(1.4, 1) circle (0.02);

			\draw[fill]	(0.6, 1.2) circle (0.02);
			\draw[fill]	(0.8, 1.2) circle (0.02);
			\draw[fill]	(1, 1.2) circle (0.02);
			\draw[fill]	(1.2, 1.2) circle (0.02);
			\draw[fill]	(1.4, 1.2) circle (0.02);

			\draw[fill]	(0.6, 1.4) circle (0.02);
			\draw[fill]	(0.8, 1.4) circle (0.02);
			\draw[fill]	(1, 1.4) circle (0.02);
			\draw[fill]	(1.2, 1.4) circle (0.02);
			\draw[fill]	(1.4, 1.4) circle (0.02);

			\draw[->]	(0, 1.2) node[left] {\scriptsize req} -- (0.6, 1.2);
			\draw[->]	(0, 1) node[left] {\scriptsize data} -- (0.6, 1);
			\draw[<-]	(0, 0.8) node[left] {\scriptsize bit ack} -- (0.6, 0.8);

			\draw[->]	(1.4, 1.2) -- (2, 1.2) node[right] {\scriptsize req} ;
			\draw[->]	(1.4, 1) -- (2, 1) node[right] {\scriptsize data} ;
			\draw[<-]	(1.4, 0.8) -- (2, 0.8) node[right] {\scriptsize bit ack} ;

			\draw[->]	(0.8, 0) -- (0.8, 0.6);
			\draw[->]	(1, 0) -- (1, 0.6);
			\draw[<-]	(1.2, 0) -- (1.2, 0.6);

			\draw[->]	(0.8, 1.4) -- (0.8, 2);
			\draw[->]	(1, 1.4) -- (1, 2);
			\draw[<-]	(1.2, 1.4) -- (1.2, 2);

		\end{tikzpicture}

		\\
		{\scriptsize (a) Chip Pins}
		&&
		{\scriptsize (b) Two inputs/outputs}
		&&
		{\scriptsize (c) Communication: Pin allocation}
	\end{tabular}
	\end{center}
	\vspace{-0.2cm}
	\caption{Controller: (a) number of controller chip pins, (b) position of two unidirectional inputs and two unidirectional outputs, (c) pin allocation for communication. \label{fig:controller}}
	\vspace{-0.3cm}
\end{figure}
The current implementation technology chosen for the controller chip has a limitation of 25 input/output
signal pins as shown in Fig. \ref{fig:controller}~(a).
The adoption of an asynchronous circuit leads to
the implementation of a four-phase asynchronous handshake
communication protocol \cite{nowick2015asynchronous} between two controllers.
The protocol requires three input/output
pins per controller to transmit a single bit. The design 
choice to accommodate these restrictions was to
allow two input channel endpoints and two output channel
endpoints as shown in  Fig. \ref{fig:controller}~(b).
This design choice leads to the allocation of 12 pins
for bit-by-bit communication (3 pins per channel endpoint)
as shown in Fig. \ref{fig:controller}~(c). The rest of the
13 pins are used for configuring the meta-material and for global signals.

\begin{figure}[t]
	\centering
		\includegraphics[width=5 cm]{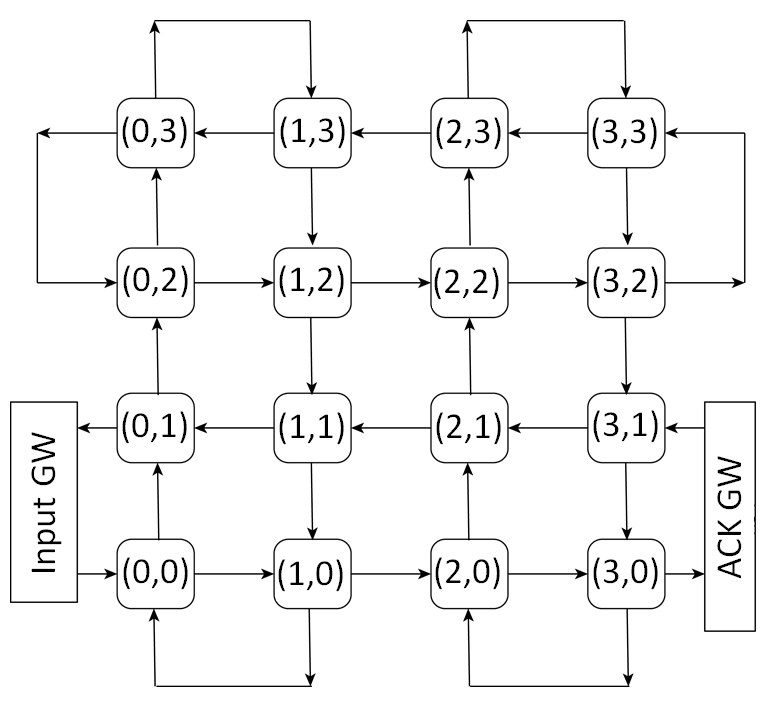}
	\vspace{-0.2cm}
	\caption{Manhattan-like network with edge wrap-around, with the input gateway shown at the south west corner and the ACK gateway shown is at the south east corner of the network.}
	\vspace{-0.2cm}
	\label{the_grid}
\end{figure}

The next design choice is the CN topology which is
shown for a $4 \times 4$ grid in Fig. \ref{the_grid}.
It is a Manhattan-like topology with no wraparounds.
The choice not to include wraparound links is justified by the
limitations on the hardware design; the printed circuit
board requires the implementation of two more layers for wraparound
circuits and, moreover, controllers at the edges
would require transistors with a higher signal drive to
achieve the sending of a signal through the wraparound. Instead, in the CN topology nodes that reside on the same edge row (column) are connected between them through \textit{edge wraparounds} as shown in Fig. \ref{the_grid}. The Manhattan-like connectivity was chosen mainly due to
the controller communication capabilities.
In addition, it is a more robust design in relation to
other topologies such as a monotonous topology, since
in the absence of wraparounds it connects every pair
of switch controllers.

The network is configured via two GWs which are also responsible for traffic generation. GWs are "smart" devices that connect the CN to external world communication and may also have environmental sensory capabilities.  
\section{Proposed Routing Schemes}
\label{algorithms}
\begin{figure}[b]
	\centering
		\includegraphics[width=7 cm]{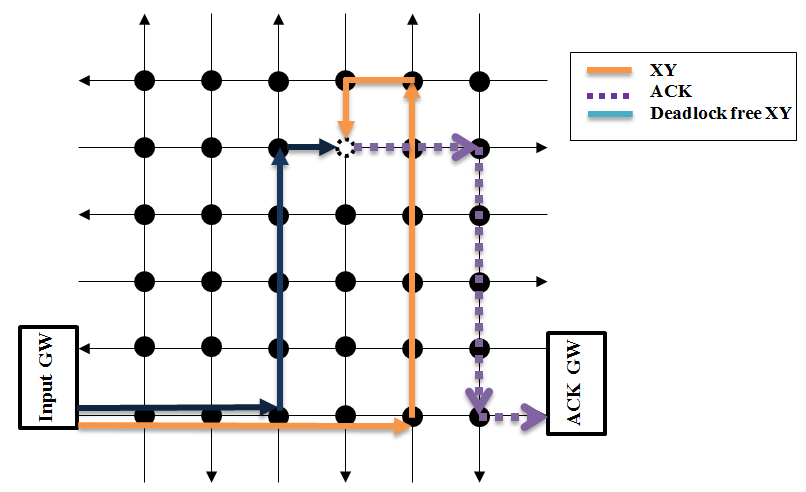}
	\vspace{-0.2cm}
	\caption{Paths taken by: (1) classic XY routing (orange), (2) deadlock-free XY (blue), and (3) the ACK packet (purple).}
	\vspace{-0.2cm}
	\label{deadfreexy}
\end{figure}

\begin{figure}
	\begin{center}
	\begin{tikzpicture}[scale=.9]

		\draw[rounded corners]	(0, 0) rectangle (0.5, 0.5);
		\node		at	(0.25, 0.25) {\scriptsize $0$};
		\draw[->]	(-0.5, 0.25) -- (0, 0.25);
		\draw[->]	(0.5, 0.25) -- (1, 0.25);
		\draw[->]	(0.25, -0.5) -- (0.25, 0);
		\draw[->]	(0.25, 0.5) -- (0.25, 1);

		\draw[rounded corners]	(1.6, 0) rectangle (2.1, 0.5);
		\node		at	(1.85, 0.25) {\scriptsize $1$};
		\draw[->]	(1.1, 0.25) -- (1.6, 0.25);
		\draw[->]	(2.1, 0.25) -- (2.6, 0.25);

		\draw[<-]	(1.85, -0.5) -- (1.85, 0);
		\draw[<-]	(1.85, 0.5) -- (1.85, 1);

		\draw[rounded corners]	(0, 1.6) rectangle (0.5, 2.1);
		\node		at	(0.25, 1.85) {\scriptsize $2$};
		\draw[<-]	(-0.5, 1.85) -- (0, 1.85);
		\draw[<-]	(0.5, 1.85) -- (1, 1.85);

		\draw[->]	(0.25, 1.1) -- (0.25, 1.6);
		\draw[->]	(0.25, 2.1) -- (0.25, 2.6);

		\draw[rounded corners]	(1.6, 1.6) rectangle (2.1, 2.1);
		\node		at	(1.85, 1.85) {\scriptsize $3$};
		\draw[<-]	(1.1, 1.85) -- (1.6, 1.85);
		\draw[<-]	(2.1, 1.85) -- (2.6, 1.85);

		\draw[<-]	(1.85, 1.1) -- (1.85, 1.6);
		\draw[<-]	(1.85, 2.1) -- (1.85, 2.6);
	
%
%

	\end{tikzpicture}
	\end{center}
	\vspace{-0.2cm}
	\caption{Controller four different orientations. \label{fig:controller_orientations}}
	\vspace{-0.2cm}
\end{figure}
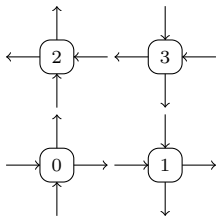

Due to its simplicity the XY algorithm has been chosen as the basis of the algorithms to be developed. 
The XY is deadlock free for 2 dimensional mesh networks as pointed out in \cite{chiu2000odd, glass1992turn}. Nonetheless, regular XY can cause deadlocks in the HSF CN because of the fact that the communication direction changes on consecutive rows and consecutive columns, as shown in Fig.\ref{the_grid}. Hence, a variant of  XY routing, which is orientation-aware can be employed to ensure deadlock freedom. Fig. \ref{deadfreexy} demonstrates the difference between classic XY and the deadlock-free XY variant proposed in the HSF CN. When the destination is on an odd column, regular XY uses the even column after the destination column to deliver the packet. However, when multiple packets are sent to the same destination, a cycle (and hence deadlock) can be created because of the ACK packet as shown in Fig.\ref{deadfreexy}. The deadlock-free XY variant on the other hand, uses the even column that is before the destination column to send upwards, which avoids creating cycles by not routing to the west, as depicted in Fig. \ref{deadfreexy}. This variant of the XY however, is not fault tolerant and offers no adaptability.
The HSF application is expected to sustain moderate to low traffic rates, which can reduce the probability of deadlocks dramatically. In addition, the limited adaptability of the network topology renders the deployment of traditional fault tolerant routing techniques infeasible. Thus, at this stage of our work we focus on loop-free routing and reliable data delivery. In this section, we propose two fault-adaptive routing techniques. 
The proposed algorithms use the orientation and coordinates of the destination node and the faulty nodes to provide fault adaptability. According to the location of the controller, there are four different types of orientations, namely, nodes on an even row and an even column (type $0$), nodes on an odd row and an even column (type $1$), nodes on an even row and an odd column (type $2$) and nodes on an odd row and an odd column (type $3$), as depicted in Fig. \ref{fig:controller_orientations}. 
\subsection{Loop-Free Algorithm (LFA)}

As noted in Section \ref{intro}, the turn model, which has been adopted in multiple works \cite{chiu2000odd, wu2003fault}, is not suitable for the network under consideration. Each router in the CN has two unidirectional output channels. Since the turn model removes at least one turn in each column to ensure deadlock freedom, many of the routers will be left with one routing direction which can significantly drop delivery ratios in the presence of faults. Classical XY routing on the other hand, is not FT. Our proposed algorithm integrates turn prevention with a fault-adaptive XY variant, denoted by XY-YX, to avoid loops and tolerate faults simultaneously. 

XY-YX employed by the loop-free algorithm (LFA) utilizes the nodes awareness of the status of their output channels to alternate between using XY and YX routing. Each node in the considered network has two output channels. A channel is considered faulty if it is defective, or in the case it is non-faulty but leads to a non-functioning node. The main idea is that the packet is initially forwarded using XY until a faulty channel is detected in its path. In this case, the packet is directed to the alternative healthy output channel of the node after altering the header, such that starting from the next node YX is to be employed. When, another fault is detected the same process is repeated but the routing technique is swapped with XY, and so on. Using MATLAB simulation experiments (not included in this paper) XY-YX indicated considerable improvement in successful delivery rates compared to other XY variants. However, an unaided XY-YX is prone to live locks in the presence of faults as the algorithm can sometimes forward the packet to the same faulty path.Thus, to avoid loops and yet maintain routing flexibility, we propose to use in the vicinity of faults based on the destination node location and orientation, a version of the turn model on top of the XY-YX. It is worth noting that not all fault-destination pairs will create a live-lock. Therefore, we limit the usage of the turn prevention policy to the locations where faults are expected to create loops. The routing mode in which turn prevention is employed is referred to as \textit{abnormal} routing, and the mode where simple XY-YX is employed is referred to as \textit{normal} routing.

The method works as follows: a packet is routed in the \textit{normal} routing mode using XY until a fault is encountered, which is when (based on the location of the fault) the packet's header is altered such that the routing algorithm is set to YX, and if necessary, the routing mode is switched to the \textit{abnormal} routing. The abnormal routing restricts certain turns, so as to break potential cycles that would cause deadlocks, and deliver the packet from the input channel that was not targeted in the previous routing mode. In other words, if the XY routing would deliver a packet through input $1$ of the destination node, YX (with the abnormal mode if necessary) targets input $2$. This will mostly route the packet around the faulty node. However, there exist special cases where a fault can leave the destination node disconnected from the network, in which case routing to destination is impossible (e.g., when node (2,2) in Fig. \ref{the_grid} is faulty, nodes (2,3), (3,2) and (3,3) become disconnected). It is important to mention here that this scheme forbids $180 ^\circ$ turns at the edge wraparounds at all times. LFA requires two bits in the header of a packet, one to indicate the routing technique, either XY or YX, and the other to indicate the routing mode. The node prior to the fault is responsible for selecting the correct routing mode which is accomplished by comparing the fault node location and orientation with the target location and orientation. In addition, having an extra bit in the header to indicate the routing mode allows the method to terminate when a loop cannot be avoided. Hence, if abnormal mode is employed and the packet encounters an additional fault that forces taking a blocked turn, the algorithm terminates to avoid live-locks. The detailed set of rules of the algorithm is not included in the paper due to space limitations. Instead, we present a walk-through example and leave the details to follow in future publications. 

\subsubsection{Walk-through Example}
Fig. \ref{routess} shows the paths taken by the different routing algorithms to the destination node (white). The destination is type $3$ which means that it lies on an odd column and an odd row. The LFA starts with XY. When the fault is detected at the node prior to the fault, each algorithm takes a different action. XY will stop and stall the packet, while LFA will forward the packet to the healthy output of the current node (send east) and then re-route starting from node $a$. At $a$, LFA switches the routing algorithm to YX which would normally forward the packet upwards to eliminate the offset in the vertical axes, and then left to eliminate the horizontal offset. Nevertheless, this will lead the packet to the same faulty node which will create a cycle. Therefore, in LFA the only turn allowed at $a$ in this case is east south (i.e. to send south). At the node below $a$, no turns are restricted, and YX will send the packet to $b$ which if not forbidden will send the packet upwards leading to the same cycle. Therefore, the blocked turn at $b$ is the west north and thus, from $b$ the packet will be forwarded west to node $c$. From $c$, YX is used to route to the destination without the need to further block turns. So, to prevent the YX method from entering loops, the LFA forbids two turns in the vicinity of the fault in this case, namely, the east north turn at the odd column which is higher than the target's column and west north at the even column.   
\vspace{-0.2cm}
\subsection{Reliable Delivery Algorithm}
This method employs different variants of the XY and YX techniques to provide two disjoint paths between every pair of nodes in the network. To ensure orthogonality, each path must target a different input channel of the destination node, and each path must be in an opposite direction of the other. In other words, one path is in the clockwise direction and the other is in the counter-clockwise direction. The reliable delivery algorithm (RDA) starts with targeting one of the input channels of the destination based on the orientation and location of the destination node. When a fault is detected the packet is rerouted towards the path that targets the other input that was not targeted at the beginning. The availability of two separate routes from any source node to any destination node in the network, offers notable flexibility in the presence of faults. 
The different paths are selected based on the location of the current (source) node and the orientation and location of the destination node. The considered topology creates a repetitive pattern which can be used to always find two disjoint paths between two nodes. The way a packet is forced to take a certain path is by blocking specific turns regardless of the existence of a faulty node.

\begin{figure}[t]
	\centering
		\includegraphics[width=7 cm]{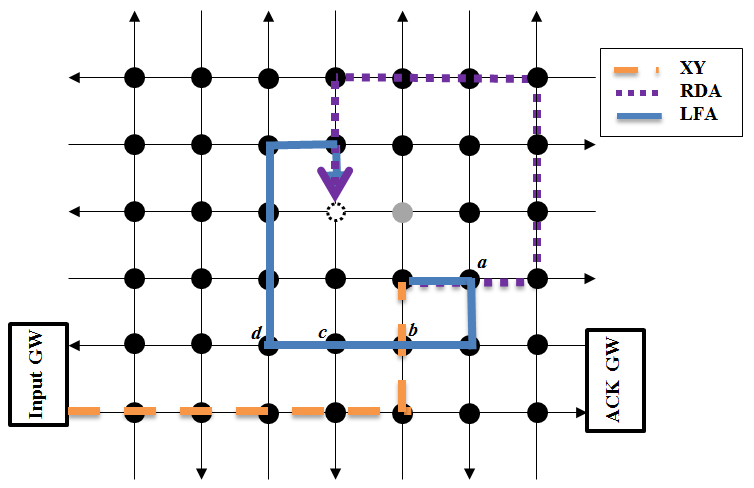}
	\caption{Routes taken from the node prior to the fault (gray) to the destination (white) by the XY, RDA and LFA.}
	\vspace{-0.2cm}	
	\label{routess}
	\vspace{-0.2cm}
\end{figure}

As the detailed rules are omitted, the example of Fig. \ref{routess} is used to clarify how the algorithms work. In Fig. \ref{routess} the destination is type 3, hence RDA starts with the same path as XY. However, when the fault is detected, RDA directs the packet to take a path that does not overlap with the previous one by blocking east turns at the even column that is after the destination until the packet is at a row higher than the destination. Then the packet is forwarded east south at the destination column.    
\section{Performance Evaluation}
\label{SimEx}
As mentioned earlier in the paper, the routers in the HSF CN have no clocks and thus employ asynchronous communication. The available NoC simulators, however, are mostly based on scheduling and do not offer a ready-to-use clock-less communication option. Therefore, using AnyLogic \cite{borshchev2013big}, we built a custom-made simulator that employs the four-way handshake, which allows routers to communicate asynchronously. Our HSF CN Asynchronous Simulator (HCNAS) relies on conditional events to achieve asynchrony. The network is connected to two "smart" gateways which are equipped with clocks and are responsible for generating data traffic and handling ACKs. A Node comprises of two output ports and two input ports, which connect each router to its neighbors, as shown in Fig. \ref{the_grid}. Edge nodes are connected to each other through edge wraparounds, which creates a bidirectional channel between each two edge neighbors. In addition, a node has enough buffering space to receive an entire packet before forwarding it. Faulty nodes are created at the initialization level before the data enters the network. Nodes perform the four-way hand shake to receive and transmit every bit, until the whole packet is processed. Once an output port (input port) is selected for transmission (reception) it cannot be changed until the current transmission (reception) of the packet is over. Also, a node cannot receive and transmit at the same time, nor receive from or transmit to two different nodes simultaneously. Based on the packet header, nodes locally decide the direction of the next hop. Furthermore, nodes have local knowledge of their faulty neighbors, and thus block the output ports connected to faulty nodes. Thus, when a packet "should" be transmitted through a faulty channel, it can either be redirected to the healthy channel, get stalled, or destroyed, based on the routing algorithm. If both outputs are blocked packets can be dropped or stalled.  
\subsection{Simulation Results}
The simulation experiments are conducted for a $24 \times 24$ Manhattan-like network with edge wraparounds, to evaluate the performance of the proposed routing algorithms in the presence of faults. The input GW and ACK GW are assumed to be at the bottom left corner and bottom right corner of the network, respectively. This is a design choice that enables employing XY and YX and ensures deadlock freedom of the XY in the considered topology. Each node has a probability of failure equal to $P_{f}$, where $P_{f}$ assumes values in the range $0.0 - 0.08$ in increments of $0.02$. This range gives sufficient information to obtain the general pattern of the performance. A single packet is sent from the input GW to the destination node. In case of successful delivery, an ACK packet is forwarded from the destination to the ACK GW. 

\begin{figure}[t]
	\centering
		\includegraphics[width=4 cm]{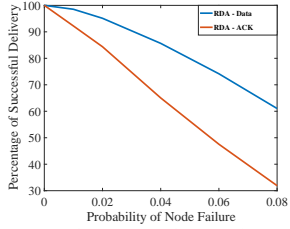}
	\vspace{-0.2cm}
	\caption{Percentage of successful delivery of data packets to the destinations and the ACK packets to the acknowledgment gateway.}
	\vspace{-0.5cm}
	\label{algb_res}
\end{figure}

Fig. \ref{algb_res} shows the overall percentage of successful delivery of data packets to the destinations and the ACK packets to the ACK gateway, for RDA. Four destinations of four different orientations are selected at each quarter of the network. For each destination node, the experiment is repeated $1000$ times for each $P_{f}$ value. The algorithm achieves more than $98\%$ successful delivery at the destination node for failure probability of $0.01$, as shown in Fig. \ref{algb_res}. However, this success percentage drops below $90\%$ when the fault probability is greater than $0.04$. The discrepancy between the delivery rates of the data packets and those of the ACK packets, is caused by the fact that the percentage of the delivered ACK packets is measured with respect to the number of data packets sent from the input GW regardless of the delivery status at the destination node. 

\begin{figure}[t]
	\centering
		\includegraphics[width=9 cm]{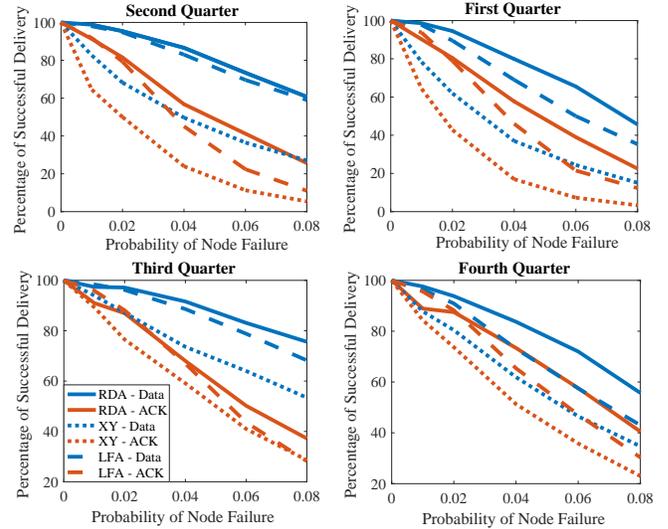}
	\vspace{-0.2cm}
	\caption{Percentage of successful delivery of data packets to destinations in the four different quarters of the network and of ACK packets to the ACK gateway, for RDA, XY and LFA.}
	\vspace{-0.2cm}
	\label{allQ}
\end{figure}

We compare the performance of the RDA, LFA with the XY in the presence of faults as shown in Fig. \ref{allQ}. A packet is sent to four destinations, each in a different quarter of the network, and the experiment is repeated $1000$ times. The four selected destination nodes are of different orientations to ensure diversity in the results. From the figure, it can be observed that RDA achieves high success rates compared to the other methods. In addition, in spite of the fact that the current version of the LFA is designed to avoid loops for one fault only, it achieves comparable results with RDA, especially, when the destination is in the second or third quarter. Overall, successful delivery ratios decline when the destination is further away from the input gateway, which is reasonable as the random fault model used implies that longer paths are more likely to have faults. 
LFA has been evaluated in the case of a single fault in the path to the destination. Four destinations in the four quarters of the network have been considered. A node in the path to the destination is randomly selected to be faulty and the experiment has been repeated $200$ times for each destination location. LFA achieves $100\%$ successful delivery rate at the destination and the ACK GW. The graphs are not included in the paper. 
\section{Conclusions}
\label{conc}
HyperSurfaces are novel planar devices that offer customizable interaction with electromagnetic waves. A core component is an embedded network of miniaturized controllers,  with unique 
restrictions in terms of fault tolerance. In this work we propose HyperSurface-specific fault adaptive techniques, based on XY variants and modified versions of the turn model. The effectiveness of the proposed schemes is shown through simulations. The principal aim in the near future is to develop a fault tolerant routing algorithm which ensures deadlock and live-lock freedom.
\bibliography{reff}

\begin{thebibliography}{10}

\bibitem{liaskos2015design}
C.~Liaskos, A.~Tsioliaridou, A.~Pitsillides, et~al.
\newblock Design and development of software defined metamaterials for
  nanonetworks.
\newblock {\em IEEE Circuits and Systems Magazine}, 15(4):12--25, 2015.

\bibitem{liaskos2018new}
C.~Liaskos, S.~Nie, A.i Tsioliaridou, A.~Pitsillides, S.~Ioannidis, and
  I.~Akyildiz.
\newblock A new wireless communication paradigm through software-controlled
  metasurfaces.
\newblock {\em IEEE Communications Magazine}, 56(9):162--169, 2018.

\bibitem{yang2016programmable}
H.~Yang, X.~Cao, F.~Yang, et~al.
\newblock A programmable metasurface with dynamic polarization, scattering and
  focusing control.
\newblock {\em Scientific reports}, 6:35692, 2016.

\bibitem{abadal2017computing}
S.~Abadal, C.~Liaskos, A.~Tsioliaridou, et~al.
\newblock Computing and communications for the software-defined metamaterial
  paradigm: A context analysis.
\newblock {\em IEEE access}, 5:6225--6235, 2017.

\bibitem{petrou2018asynchronous}
L.~Petrou, P.~Karousios, and J.~Georgiou.
\newblock Asynchronous circuits as an enabler of scalable and programmable
  metasurfaces.
\newblock In {\em proceedings of the ISCAS}, pages 1--5. IEEE, 2018.

\bibitem{kouvarosformal}
P.~Kouvaros, D.~Kouzapas, A.~Philippou, et~al.
\newblock Formal verification of a programmable hypersurface - work in
  progress.
\newblock In {\em Proceedings of the 23rd FMICS}, pages 83--97, 2018.

\bibitem{vangal200880}
S.~R. Vangal, J.~Howard, G.~Ruhl, et~al.
\newblock An 80-tile sub-100-w teraflops processor in 65-nm cmos.
\newblock {\em IEEE Journal of Solid-State Circuits}, 43(1):29--41, 2008.

\bibitem{radetzki2013methods}
M.~Radetzki, C.~Feng, X.~Zhao, et~al.
\newblock Methods for fault tolerance in networks-on-chip.
\newblock {\em ACM Computing Surveys}, 46(1):1--38, 2013.

\bibitem{glass1992turn}
C.~Glass and L.~Ni.
\newblock The turn model for adaptive routing.
\newblock {\em ACM SIGARCH Computer Architecture News}, 20(2):278--287, 1992.

\bibitem{vitkovskiy2013dynamic}
A.~Vitkovskiy, V.~Soteriou, and C.~Nicopoulos.
\newblock Dynamic fault-tolerant routing algorithm for networks-on-chip based
  on localised detouring paths.
\newblock {\em IET Computers \& Digital Techniques}, 7(2):93--103, 2013.

\bibitem{aisopos2011ariadne}
K.~Aisopos, A.~DeOrio, L.~Peh, et~al.
\newblock Ariadne: Agnostic reconfiguration in a disconnected network
  environment.
\newblock In {\em Proceedings of the PACT}, pages 298--309. IEEE, 2011.

\bibitem{wu2003fault}
J.~Wu.
\newblock A fault-tolerant and deadlock-free routing protocol in 2d meshes
  based on odd-even turn model.
\newblock {\em IEEE Transactions on Computers}, 52(9):1154--1169, 2003.

\bibitem{nowick2015asynchronous}
S.~Nowick and S.~Montek.
\newblock Asynchronous design—part 1: Overview and recent advances.
\newblock {\em IEEE Design \& Test}, 32(3):5--18, 2015.

\bibitem{chiu2000odd}
G.~Chiu.
\newblock The odd-even turn model for adaptive routing.
\newblock {\em IEEE Transactions on parallel and distributed systems},
  11(7):729--738, 2000.

\bibitem{borshchev2013big}
A.~Borshchev.
\newblock {\em The big book of simulation modeling: multimethod modeling with
  AnyLogic 6}.
\newblock AnyLogic North America Chicago, 2013.

\end{thebibliography}

\end{document}